\documentstyle[preprint,aps]{revtex}
\begin{document}
\draft
\date{\today}

\title{PRESSURE INDUCED HYDRATION DYNAMICS OF MEMBRANES}

\author{F. \"{O}sterberg, M. Kriechbaum\cite{a}, A. Polcyn\cite{b}, V.
Skita\cite{c}, M. W. Tate,
P. T. C. So\cite{*}, S. M. Gruner, and
Shyamsunder Erramilli \cite{footnote}}
\address{
Department of Physics, Joseph Henry Laboratories
Princeton University
P.O. Box 708
Princeton, NJ  08544
}

\maketitle
\begin{abstract}

	Pressure-jump initiated time-resolved x-ray diffraction studies
of dynamics of the hydration of the hexagonal phase in
biological membranes show
that (i) the relaxation of the unit cell spacing
is non-exponential in time;  (ii) the Bragg peaks shift
smoothly to their final positions without significant broadening or loss in
crystalline order. This suggests that the hydration is not diffusion
limited
but occurs via a rather homogeneous swelling of the whole lattice,
described by power law kinetics with an exponent $
\beta = 1.3 \pm 0.2$.

\end{abstract}


\narrowtext

The availability of high brightness x-ray synchrotron sources
and a new generation of detectors that
acquire complete 2-D Bragg diffraction patterns at video rates
\cite{gruner87,caffrey} enables the
direct structural study of the  dynamics of lyotropic liquid
crystalline systems  on millisecond time-scales.
This allows  study of diffusion and water transport in these
topologically complex systems which are of interest not only to the
biophysics community, but also generally to those interested in fluid
invasion into porous media with nanometer sized pores.

Dispersions of biological lipids in water exhibit a rich variety of
liquid crystalline phases in which membranes form periodic lattices in
either 1, 2 or 3 dimensions\cite{tate91}. The
inverted hexagonal ($H_{II}$) phase (Fig. 1a, inset)
is one of the
simpler non-lamellar phases and has relevance to biological processes
as well\cite{gruner85}. This $H_{II}$ phase is formed by a
periodic lattice of cylindrical pores of water $\sim 40$ \AA\ in
diameter, each surrounded by a monolayer of lipid molecules.
Detailed studies\cite{dope_references} done on such
systems in equilibrium have shown \cite{kirk84,charvolin85} that
the thermodynamics of such non-lamellar phases can be understood
as a result of a {\it frustration} between the opposing requirements that
membranes have to curl to satisfy curvature requirements on the one hand
but also to maintain constant bilayer thickness on the
other.
This frustration results in extreme sensitivity of the $H_{II}$ unit cell
spacing $a$ to changes in both temperature and pressure \cite{so93}.
Application of hydrostatic pressure in the presence of excess
water \cite{excess_water}
leads to an {\it increase} in $a$, due primarily to an increase in the
diameter of the water core, and hence, in the
number of water molecules $n_{w}$ associated with each lipid molecule
\cite{so93,turner}.
A pressure jump can therefore be used to  induce a sudden change in the
chemical
potential of water in order to drive water into or out of
the $H_{II}$ domains. The dynamics of water uptake/release can then be
examined.

{\it Time-resolved studies:}
Time-resolved x-ray diffraction studies of the $H_{II}$ phase were done at
beamline X9A at the Brookhaven National Synchrotron Light Source, using
pressure jumps in the range 0.1-2 kbar over a
temperature range of 280-350 K.
The beam had an intensity of about $10^{11}$ photons/sec in a $\sim 500 \mu$m
spot, and an energy of $E=8$ keV, with a resolution of $\Delta E/E
\sim 10^{-4}$.
Samples of about 50 mg of material were contained in a
temperature-controlled high-pressure cell
with beryllium windows.
Unit cell spacings were obtained from
positions of the Bragg peaks with calibration to silver stearate.
The instrumental resolution (FWHM)
for the scattering angle was $\Delta \theta \sim 1\times 10^{-3}$ rad.
Fast pressure jumps with a rise-time $\sim 7$ ms were
obtained by the opening of a high-pressure valve driven by a
computer-actuated pneumatic piston, similar to the method of
Caffrey and co-workers \cite{caffrey93}. The pressure jumps
resulted in small temperature changes
($\sim 2$ K for a 0.5 kbar jump) as measured by a fast response thermocouple
embedded within the sample.
A home-built
intensified x-ray detector system based on a
2-dimensional video rate CCD imager running in non-interlaced mode
(with a frame
time of 16 ms) was used to obtain more than one million x-ray diffraction
patterns in order to investigate both pressure-induced phase transitions
(to be reported elsewhere),
as well as the pressure-induced changes in the structure of a single
phase reported in this letter.

{\it Sample preparation:}
$H_{II}$ phases  formed by aqueous dispersions of the lipid DOPE \cite{dope}
were
investigated, both with and without addition of a small amount ($\sim
0.05$ w/w) of dodecane. The presence of the alkane extends the $H_{II}$
phase boundary permitting studies over a much larger pressure range
\cite{gruner85}. All experiments were performed on hydrated ``powder''
samples, which contain a large number of randomly oriented crystallites
of several microns size.  Samples were prepared by mechanically
mixing $\sim 200$ mg of lipid (or lipid+alkane) with an equal amount of
water.  Even with mechanical mixing, full hydration of the liquid
crystalline domains can take many days. Samples that were not fully
hydrated showed broad Bragg diffraction peaks, indicative of a wide
distribution in water concentration caused by the slow diffusion of
water through many layers of lipid.  Hydration was sped by incubating
the sample at 10 C and 1500 bar for several hours. Subsequent
experiments were performed at higher temperatures and lower pressures
which require a lower concentration of water for full hydration of the
liquid crystal.  Fully hydrated samples exhibited sharp Bragg
diffraction patterns and were used for the pressure jump studies
reported here.  Samples were checked for degradation by thin-layer
chromatography both before and after x-ray exposure.

{\it Results:} Figure 1 shows diffraction from a aqueous DOPE-dodecane
sample at 318\,K undergoing a pressure jump from 100 to 820 bar.  Fig
1a shows a representative plot of scattered intensity as a function of
the magnitude, $q$, of the scattering vector. The diffraction peaks can
be fit to Lorentzian profiles with the peak positions yielding the
spacing, $a$, of the hexagonal lattice.  Several striking aspects of
the data are illustrated in Figs. 1-2.
(i)
The peak positions relax continuously to their final value without
significant increase in disorder.  Note that the peak widths in any
given frame are seen to increase with increasing diffraction order,
characteristic of samples in which Bragg peaks are broadened by static
lattice disorder (``disorder of the second kind'') rather than thermal
(Debye-Waller) effects \cite{guinier}.  Analysis of the Lorentzian
profile of the peaks, correcting for the finite beam width using
standard techniques \cite{guinier}, shows that the density-density
correlation function decays as $<\rho (0)\rho(r)> \sim e^{-r/\xi}$ with
$\xi =1800 \pm 500 \AA$.  During the pressure jump shown in Fig. 1,
the angular peak width $\Delta \theta$ of the (10) peak increases by
$\sim 6 \times 10^{-5}$ rad,
much less than shift in the peak position of $2 \times 10^{-3}$ rad
from its initial to final value.
(ii)
Immediately following the pressure jump, the relative intensities of
the diffraction orders change, indicating a change in molecular
arrangement has occurred. These intensity changes are consistent with
those observed in equilibrium studies at various hydrations and
pressure, suggesting a quasi-static picture in which elastic
deformations have relaxed on time-scales shorter than 10 ms.  On a
longer time scale (seconds), the liquid crystal adjusts its composition
(i.e. $n_w$) in response to the change in the chemical potential of
water brought on by the change in pressure. This change in composition
is manifest by the change in spacing, $a$.
(iii)
The relaxation of the lattice spacing, $a$, is non-exponential in time
(Figure 2a,b). Moreover, the data from all the pressure jumps collapse
on to a single curve if they are rescaled by a time constant
and by an amplitude transform of the form
\begin{equation}
\Delta a_{norm} = \frac{a(t)-a(0)}{a(\infty)-a(0)},
\end{equation}
where $a(0)$ and $a(\infty)$ are the initial and final lattice spacings
which depend only on the initial and final pressures of the jump sequence
(Figure 2b).

Given the quasi-static picture, the measurement of $a$ as a function of
time, $t$, can be combined with equilibrium high-pressure dilatometric
data\cite{so93} to measure the number of water molecules per lipid molecule,
$n_{w}(t)$ (Fig. 2d).
$n_{w}$ is related to $a$ and the radius of the water core, $R_{w}$,
by a simple geometric relation:
\begin{equation}
n_{w}=\frac{v_{l}}{v_{w}}\frac{2\pi R_{w}^2}{\sqrt{3} a^{2} - 2\pi R_{w}^{2}}.
\end{equation}
The ratio $v_{l}/v_{w}$ of the molecular volumes of the lipid and
water is known from high-pressure dilatometry measurements
\cite{so93}, and shows that the increase in $a$ under pressure is due
to an increase in $n_{w}$. $R_w(a,P)$ was found by static x-ray
measurements\cite{so93}.  Measurements of the intrinsic crystalline
disorder $\Delta a$ gives us the variance $\Delta n_{w}$. The quality
of our data are such that only the first and second moments of the
probability distribution $p(n_{w},t)$ can be obtained. Fig 3 shows a
plot of $p(n_{w},t)$ as a function of time, and shows that the mean
value, $\overline{n_{w}}$, also relaxes non-exponentially (Fig 2c) while the
width of the distribution $p(n_{w},t)$ does not change significantly.
We may assume that, to a first approximation, the system may be
described just by specifying the mean number $\overline{n_{w}}(t)$ as a
function of time. We define a rescaled relaxation function
\begin{equation}
\zeta(t)=\frac{\overline{n_{w}}(t)-\overline{n_{w}}(\infty)}{\overline{n_{w}}(0)-\overline{n_{w}}(\infty)}.
\end{equation}
$\zeta(t)$ relaxes from an initial value of 1 to a final value of 0.
Non-exponential kinetics are often analyzed empirically by fitting the
observed decays to either stretched exponential or power laws,
particularly in the long time limit.  We find that the data may be fit
to the empirical form $\zeta(t) = (1+t/\tau)^{-1/\beta}$.  A plot of
$(\zeta^{-\beta}-1)$ as a function of time appears linear over nearly
three decades in time (Fig 2d) \cite{caveat}, as can be seen in Fig.
2d.  All pressure jumps, both upwards and downwards, can be fit to this
functional form with an exponent of $\beta = 1.3 \pm 0.2$.  For a
sample of a given chemical composition, the characteristic
time-constant depends on temperature, but is relatively insensitive to
the pressure or even the direction of the pressure jumps.  In the
absence of a microscopic description, the proposed power law decay can
only be regarded as a phenomenological characterization of the data,
but may serve as a guide to future theoretical work.
Studies at shorter times and over larger pressure ranges are needed to
improve the error bars in the exponent $\beta$, and to observe if there
are systematic deviations from the scaled collapse in Fig. 2b.

We know of no simple model which explains observations (i)-(iii).
A straightforward  explanation for the observed
non-exponential nature of the kinetics is to postulate that the observed
relaxation is due to domains of different sizes relaxing with different time
constants to the final value of the lattice spacing.  Perhaps the
simplest model is a two-state model, with the system relaxing from
its initial state to the final state. Such a model would give rise to
a diffraction pattern in which the scattering pattern at intermediate times is
a superposition of initial and final state Bragg patterns. Inspection
of Fig 1 shows that the such a superposition is not observed.
Extensions of this idea in which each domain relaxes exponentially but
independently (`in parallel') also leads to the prediction that the peak widths
following the pressure jump would be much broader than observed.

Power law kinetics arise generically in diffusion problems.
Water diffuses in from the outer
surface of the crystallite driven by the pressure-induced chemical potential
difference. If the observed kinetics are indeed determined by
diffusion of water inside the liquid crystal,
one would predict that density-density correlation would
be reduced, leading to an increase in peak-widths.
Simulations of such models suggest
that the
widths of the Bragg peaks should increase by an amount comparable to
the shifts in peak position. However this is not what is observed in
Figs 1-2.
Sophisticated treatments of concentration changes following a
perturbation have been based on dynamics of the type first analyzed by Cahn
and Hilliard (for both conserved and unconserved order parameters)
\cite{cahn60}.
Linearized versions of such treatments invariably lead to exponential kinetics.
There have
been recent reports of extensions to Cahn-Hilliard theory to take into
account non-linear effects due because the mobility may depend on the
local concentration or on time\cite{sciortino93}, but applications of this
approach
to the present problem are unclear at present. The
 phenomenological characterization of our data presented above may provide a
guide to future theoretical work.

One class of models that is consistent with the spirit of the
discussion above
are {\it sequential models} of relaxation
inspired by
spin glass literature, in which water
molecules encounter barriers of ever increasing height, with the
barrier height being renormalized by the number of water molecules
already transported.
This model would lead to a swelling of the lattice
and is consistent with the observed x-ray diffraction data. Non-exponential
kinetics
then arise from a distribution of barriers to the entry of water molecules.
Such a sequential model assumes
that diffusion of water molecules within the crystallites is not the
rate limiting process. A uniform swelling of all the cylindrical pores
is possible only
if all the  water cylinders locally have ready access to
water. This picture may be difficult to reconcile with
electron micrographs of compact
$H_{II}$ phase crystallites \cite{grunerem}. It is possible that the
crystallites observed in electron microscopy may be obtained under
rather special conditions, not typical of those in our pressure-jump
cells. However, until this question is settled, sequential models of
relaxation are not consistent with all that is known about the
$H_{II}$ phase.

The most interesting question concerns the microscopic origin of
the energy barriers involved in the non-exponential kinetics, and an
understanding of the processes determining the time-scales involved in
the transport of water within and into liquid crystalline domains.
Studies on single crystals would be useful in that the domain size effects
can be sorted out, but it is difficult to grow single crystals of the
$H_{II}$ phase. Studies of transport in different liquid crystalline
toplogies, such as the lamellar or cubic phases, may also provide
insight into the microscopic origin of the energy barriers.
Whatever the microscopic mechanism, the fact that the pores are formed by
elastic membranes  may be expected to
give rise to
nonlinear effects  rather different from conventional
studies on fluid invasion into porous media, where the substrate
forms a rigid framework.

\narrowtext

\section{ACKNOWLEDGEMENTS}

This work was supported by the ONR (Contract N00014-86-K-0396 P0001), and
by the DOE (Grant DE-FG02-87ER60522) and by the NIH (Grant
GM32614).  M. K. was supported by the Max-Kade Foundation. We
thank our colleagues  M. Aizenman, J. T. Gleeson, R. E. Goldstein, P.
Harper and O. Narayan for useful discussions.

\begin{figure}
\caption{
a) The figure shows a fit to the integrated diffraction pattern.
Fitting the Bragg diffraciton peaks to Lorentzian line-shapes, allowed
for determination of the positions, peak widths and areas of the
peaks as a function of time.
For the pattern indicated, the inter-cylinder spacing $a = 69.1$ \AA; analysis
of the
peak widths yields a variance $\Delta a = 0.6 \pm .2$ \AA which
characterizes the crystalline disorder. b) Bragg peak positions as a function
of time. Integrations
along the equatorial plane of the X-ray data obtained from time-resolved x-ray
diffraction.  An upward pressure
jump from 100 bar to 1200 bar at a temperature of $318 \pm 4$ K
occurs about two seconds after the x-ray shutter is opened.
}
\label{fig1}
\end{figure}

\begin{figure}
\caption{  a) Unit cell spacing in the $H_{II}$ phase as a
function of time.
b) Rescaled plot of $\Delta a_{norm}$ as a function of time
showing the non-exponential nature of the decay. Data shown are
from five different pressure jumps ranging in amplitude from 0.7
kbar-1.3 kbar, with changes in the unit cell spacing ranging from
$2 \AA - 7 \AA$ and appear to collapse to a single normalized curve.
The time axis has been rescaled with a characteristic
time constant $\tau$ obtained from a fit shown in Fig. 2d (see
below).
c) Observed peak width as function of time for the (1,0) and (1,1) peaks of
the inverted
hexagonal lattice.
d) A log-log plot of the normalized relaxation function
shows that the relaxation is strongly non-exponential. Solid line is a fit to a
power law form discussed in the text.}
\label{fig2}
\end{figure}

\begin{figure}
\caption{ A time sequence of the probability distribution $p(n_{w},t)$
obtained from the measured first and second moments for a
pressure jump from 1 kbar to 0.13 kbar. Curves at $240$ ms
interval have been displaced vertically for clarity. Note that the increase in
the width of the distribution is about ten times smaller than the shift in the
average position.}
\label{fig3}
\end{figure}


\begin{references}
\bibitem[a]{a} Present address: Institute f\"{u}r Biophysik and
R\"{o}ntgenstrukturforschung, Austrian Academy of Sciences,
Graz, Austria.
\bibitem[b]{b} Present address: Physics dept., Mass. Inst. of Tech., Boston,
MA-02139.
\bibitem[c]{c} Present address: University of Connecticut,
Stoors, CT.
\bibitem[*]{*} Present address: Physics dept., University of Illinois, Urbana
IL-61801.
\bibitem{footnote} To whom correspondence should be addressed.
\bibitem{gruner87} S. M. Gruner, Science {\bf 238}, 305-312 (1987).
\bibitem{caffrey} M. Caffrey, Biochemistry {\bf
26}, 6349 (1987).
\bibitem{tate91}  M. W. Tate,
E. F. Eikenberry,  D. C. Turner, E. Shyamsunder and
S. M. Gruner, Chem. Phys. Lipids. {\bf 57}, 147-164 (1991).
\bibitem{dope} Dioleoyl-Phosphatidyl-Ethanolamine
\bibitem{gruner85} S. M. Gruner, Proc. Natl. Acad. Sci. {\bf 82}, 3665-3669
(1985).
\bibitem{dope_references} Some recent reviews of non-bilayer phases are:
S. M. Gruner, J. Phys. Chem. {\bf 93}, 7562 (1989);
J. M. Seddon, Biochim. Biophys. Acta{\bf 1031}, 1-69 (1990);
 G. Lindblom and L. Rilfors, Biochim. Biophys. Acta{\bf
988}, 221-256 (1989);
K. Larsson, J. Phys. Chem. {\bf 93}, 7304 (1989).
\bibitem{kirk84} G. L. Kirk, S. M. Gruner and D. L. Stein,
Biochem. {\bf 26}, 6592-6598 (1984).
\bibitem{charvolin85} J. Charvolin J. Physique {\bf 46}, C3-173 to C3-190
(1985).
\bibitem{so93} P. T. C. So, S. M. Gruner and Shyamsunder
Erramilli, Phys. Rev. Lett., {\bf 70}, 3455-3458 (1993).
\bibitem{excess_water} `Excess water' is defined as the point at
which water added to the lipid sample does not get absorbed by
the $H_{II}$ lattice, but instead forms pools of bulk water. As
the temperature and pressure are varied, the chemical potential
of water changes leading to a change in the excess water point.
\bibitem{turner} D. C. Turner and S. M. Gruner, Biochem. {\bf 31}, 1340-1355
(1992).
\bibitem{caffrey93} A. P. Mencke, A. Cheng and M. Caffrey, Rev.
Sci. Instr. {\bf 64}, 383-389 (1993).
\bibitem{so92} P. T. C. So, S. M. Gruner and Erramilli
Shyamsunder, Rev. Sci. Instr. {\bf 63} 1763 (1992).
\bibitem{guinier} A. Guinier in {\it X-ray Diffraction} W. H. Freeman
and Company, (1963).
\bibitem{caveat} Other functional
forms may provide qualitatively good fits. Indeed, long
experience with debates between advocates of  power laws vs
stretched exponentials to describe the decay kinetics in the glass transition
literature suggests that this functional form should be regarded only
as a characterization of the non-exponential nature of the relaxation.
\bibitem{NMR} G. Lindblom and H. Wennerstr\"{o}m, Biophysical
Chemistry, {\bf 6}, 167-171 (1977).
\bibitem{cahn60} J. W. Cahn and J. E. Hilliard, J. Chem. Phys.
{\bf 31}, 688-699 (1959).
\bibitem{sciortino93} F. Sciortino, R. Bhansil, H. E. Stanley and P.
Alstrom, Phys. Rev. E {\bf 47}, 4615-4618.
\bibitem{grunerem} S. M. Gruner, K. J. Rothschild, W. J. DeGrip and N. A.
Clark, J. Physique {\bf 46} 193 (1985).

\end{references}
\end{document}